\documentclass [5p, number] {elsarticle}
\usepackage{graphicx}
\usepackage{setspace}
\usepackage{lineno}

\title {Development of a compact E$\times$B 			
microchannel plate detector for beam imaging}
\date{\today}

\author{B. B. Wiggins}
\author{Varinderjit Singh}
\author{J. Vadas}
\author{J. Huston}
\author{T. K. Steinbach}
\author{S. Hudan}
\author{and R. T. deSouza \corref{cor1}}

\address{Department of Chemistry and Center for Exploration of Energy and Matter, \\ 
Indiana University, 2401 N. Milo B. Sampson Ln, Bloomington, Indiana 47408, USA}

\cortext[cor1]{desouza@indiana.edu}

\begin{document}
\begin{abstract}

A beam imaging detector was developed by coupling a multi-strip anode with delay line readout to an E$\times$B microchannel plate (MCP) detector. This detector is capable of measuring the incident position of the beam particles in one-dimension. To assess the spatial resolution, the detector was illuminated by an $\alpha$-source with an intervening mask that consists of a series of precisely-machined slits. The
measured spatial resolution was 520$\mu$m FWHM, which was improved to 413$\mu$m FWHM by performing an FFT of the signals, rejecting spurious signals on the delay line, and requiring a minimum signal amplitude. This measured spatial resolution of 413$\mu$m FWHM corresponds to an intrinsic 
resolution of 334$\mu$m FWHM when the effect of the finite slit width is de-convoluted. To understand the measured resolution, the performance of the detector is simulated with the ion-trajectory code SIMION.

\end{abstract}

\begin{keyword}
microchannel plate detector \sep beam imaging \sep tracking detector \sep position-sensitive microchannel plate detector
\end{keyword}

\maketitle

\section{Introduction}

A new generation of radioactive beam facilities provide unique opportunities to investigate nuclei far from $\beta$-stability. However, the beam intensity
of the most N/Z exotic nuclei is typically less than 1000 ions/s posing significant challenges in imaging these beams. In the case of low energy beams, it is particularly important that the imaging detector 
introduce the least amount of material into the beam path in order to minimally distort the beam. In addition, as most accelerator
facilities are pulsed it is beneficial 
if the imaging detector has good timing characteristics. Due to their high gain, 
fast temporal response, sensitivity to a single electron, and compact size, microchannel plates (MCPs) are often used as an electron amplifier for these imaging detectors \cite{Wiza79}. 

There are several methods for providing position sensitivity with an MCP detector including: multi-strip anode \cite{Fraser84}, helical delay line 
\cite{Sobottka88, Hong16}, cross-strip anode \cite{Siegmund09}, induced signal \cite{deSouza12,deSouza15}, resistive anode \cite{Lampton79,Wiggins15,Siwal15},
and Timepix CMOS readout \cite{Tremsin11}. 
To realize a beam imaging detector
requires transport of electrons produced at a secondary-emission foil onto the surface of 
the position sensitive MCP detector situated away from the beam axis.
In one approach, a clever magnetic field arrangement provided transport of the electrons on helical trajectories onto the 
the surface of a MCP detector  \cite{Shapira00, Matos12, Rogers15}. This technique resulted in a spatial resolution 
of 588$\mu$m FWHM \cite{Matos12}. The most serious limitation of this approach is the large space occupied by this detector making its use prohibitive in many experiments.

A beam timing detector which is compact and introduces a minimal amount of material into the beam path is an E$\times$B detector \cite{Bowman78, Kraus88, Odenweller82, Steinbach14}.
Such a detector has been used to measure the time-of-flight of beam particles and reaction products in nuclear reaction studies 
\cite{O18Steinbach14,SteinbachThesis,Singh17}. 
To make the MCP in an E$\times$B detector position-sensitive we employed a multi-strip anode with delay line readout, 
which is a particularly appealing because of its simplicity and low cost. 
Moreover, due to the fast time response of the detector it is capable of resolving two particles that arrive simultaneously but are spatially separated. 
Two principal factors influence one's ability to accurately image the beam: the impact of electron transport from the electron-emission foil to the MCP  and the inherent spatial resolution
of the position-sensitive element.
In this article, we describe the design, development, and performance of an E$\times$B position-sensitive
detector suitable for imaging low-intensity radioactive beams. We explore the impact of the electron transport for this 
detector geometry on the measured resolution using the ion trajectory code SIMION \cite{SIMION}.

\section{Experimental Setup}

\begin{figure}
\begin{center}
\includegraphics[scale=0.45]{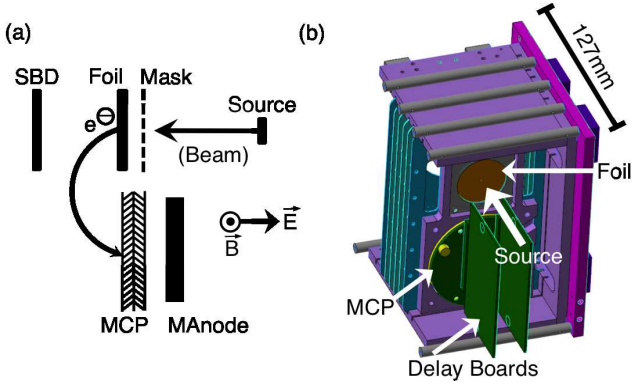}
\caption{(Color online) (a) Schematic of the experimental setup used to assess the spatial resolution of an E$\times$B MCP detector with a multi-strip anode.
(b) CAD drawing of the position-sensitive E$\times$B MCP detector. For clarity, some of the magnets along with one iron plate and the side of the PEEK box have been removed. 
} \label{fig:DSMCP}
\end{center}
\end{figure}

Presented in Fig.~\ref{fig:DSMCP}a is a schematic drawing of the experimental setup
used to determine the spatial resolution of the position-sensitive E$\times$B MCP detector. Electrons, ejected from the 0.5$\mu$m thick aluminized mylar foil by the passage of ionizing radiation, are
accelerated and bent onto the surface of a 40mm diameter MCP. The MCP used was a standard chevron stack (APD 2 MA 40/12/10/12 60:1 NR) with 10$\mu$m diameter microchannels provided by 
Photonis USA \cite{PHOTONIS}. The MCP amplifies the incident electrons by a factor of $\sim$ 10$^{6}$. The resulting electron pulse is incident 
on a multi-strip anode. Printed on an FR4 PCB, the multi-strip anode is composed of 250$\mu$m wide strips with a 75$\mu$m 
inter-strip isolation. The total active area of the anode is $\sim$ 3cm 
x 3cm (w x h), thus restricting detection of the electrons amplified by the MCP. This choice of a reduced size anode was simply due to ease of implementation in an existing setup. It served the purpose of demonstrating the feasibility of the technique. 
All 100 strips of the anode are read out 
by two independent delay boards (as indicated in Fig.~\ref{fig:DSMCP}b) to read out the even and odd strips. The use of two delay boards minimizes the attenuation and dispersion experienced by the signal in the delay line. The delay boards consist of a continuous 7771mm long trace on a 10 layer Rogers 4350 PCB with $\sim$ 1ns/tap. Construction of the delay line with a high quality PCB material is essential to minimize signal degradation. 
The position of the incident particle is measured by constructing the time difference of the signal arrival at
each end of the delay line. 
In prior work we used a multi-strip anode coupled to a delay line with a simple electrostatic arrangement 
\cite{deSouza12,Wiggins15} to achieve 
a spatial resolution of 94$\mu$m FWHM \cite{Wiggins17}. Any variation in the electron transport can only adversely impact this resolution. This delay line approach has been successfully employed at rates up to 10MHz \cite{Mizogawa97}.

A CAD drawing of the detector is presented in Fig.~\ref{fig:DSMCP}b. The electric field is produced using a series of rings situated co-axially along the beam path. 
By applying
a voltage of +4500V to the most upstream ring plate (with $\sim$ 8mm between ring plates)
and stepping the voltage down using 500M$\Omega$ resistors between each ring an electric 
field of $\sim$ 114,300 V/m is generated. A magnetic field perpendicular to the beam axis is produced by a set of 8 neodymium permanent 
magnets \cite{Magnets}. The 
magnets each measure 25mm x 25mm x 12mm and are located on two soft iron plates 
measuring 146mm x 127mm x 98mm, and collectively produce a field of $\sim$ 90G in the region 
of electron production. This magnetic field bends the path of the electrons $\sim$ 180$^{\circ}$ onto the front face of the MCP. The impact energy of the electron on the MCP is controlled by biasing the aluminized mylar foil to -1250V. The biases of the foil and the ring plates were chosen to optimize the measured 
resolution. The front of the chevron MCP was held at ground, while the back of the chevron MCP was
biased to +1792V. The multi-strip anode, spaced from the back of the MCP by 1mm, was biased to +2350V.

\begin{figure}
\begin{center}
\includegraphics[scale=0.6]{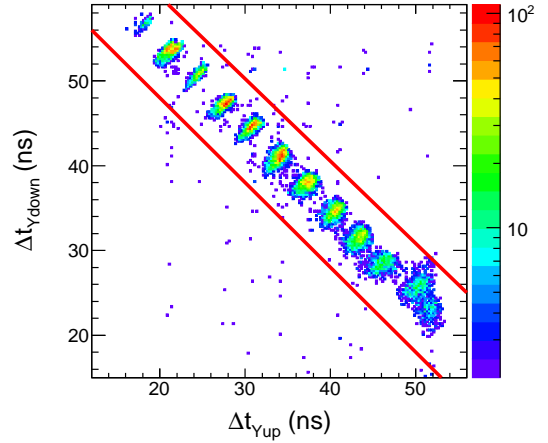}
\caption{(Color online) Two-dimensional spectrum of $\Delta$t$_{Ydown}$ $\it{vs.}$ $\Delta$t$_{Yup}$.} \label{fig:TCorrel}
\end{center}
\end{figure}

\section{Measuring the Spatial Resolution of the E$\times$B MCP Detector}

To test the performance of the detector, it was placed in a vacuum chamber that was evacuated to a pressure of 
$\sim$ 4 x 10$^{-7}$ torr and illuminated by a 1.5$\mu$Ci $^{241}$Am $\alpha$-source. Between the $\alpha$-source and the 
secondary-emission foil was a 0.8mm thick aluminum plate
with 355$\mu$m wide slits that are 6.4mm long. The 13 slits in the mask have a center-to-center spacing of 2mm. Alpha-particles passing through the mask and foil were detected using a silicon surface barrier detector (SBD) as shown in Fig.~\ref{fig:DSMCP}a.

Passage of an $\alpha$-particle through the foil generates electrons, which are transported by the E$\times$B field onto the MCP. The amplified electron signal from the MCP is incident on the multi-strip anode. A signal arriving on a strip propagates to the delay line, where it splits. The signals arriving at either end of the delay line are designated
Y$_{up}$ and Y$_{down}$, and are used to determine the position of the incident particle.
Each of these signals was amplified by a fast-timing preamplifier with a gain of 200 (Ortec VT120A) before being digitized by a 10GS/s
waveform digitizer with 8 bit resolution (Tektronix DPO5204 oscilloscope). The digitizer was triggered using a coincidence between the MCP signal and the SBD signal 
in order to reduce background events due to radiogenic decays in the MCP.
The MCP signal used for the trigger was first inverted with a 100MHz inverting transformer and subsequently amplified by an Ortec VT120A. The SBD (Ortec BA-45-900-100) was amplified by a fast preamplifier \cite{deSouza11}. Both the MCP and SBD signals were discriminated using a constant-fraction discriminator (Tennelec TC454) before forming the coincidence.

The arrival time of the Y$_{up}$ and Y$_{down}$ signals is determined by
utilizing a software constant-fraction discriminator (CFD), with a fraction of 0.5. A delay time of 1.5ns was chosen for the CFD based on the 
typical 3ns risetime of the signals. The time difference 
between the trigger time and the zero-crossing point of the CFD for the delay-line signals are designated $\Delta$t$_{Yup}$ and $\Delta$t$_{Ydown}$.

\begin{figure}
\begin{center}
\includegraphics[scale=0.50]{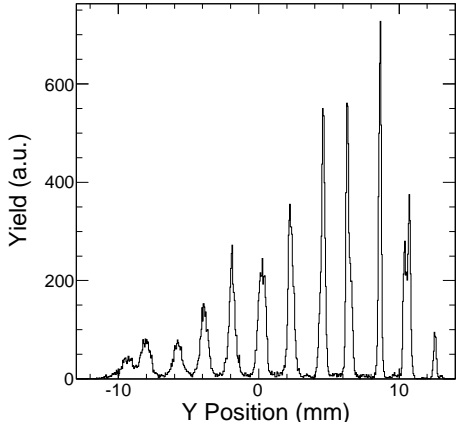}
\caption{One-dimensional position spectrum of electrons on the MCP generated by $\alpha$
particles incident on the foil that have passed through the calibration mask. 
Slits in the mask, with a width of 355 $\mu$m, have a 2mm center-to-center spacing.} \label{fig:YPos}
\end{center}
\end{figure}

The two-dimensional correlation between $\Delta$t$_{Yup}$ and $\Delta$t$_{Ydown}$
is shown in Fig.~\ref{fig:TCorrel}.
The majority of the data in Fig.~\ref{fig:TCorrel}
lies in a single anti-correlated band, with $\Delta$t$_{Yup}$ increasing as $\Delta$t$_{Ydown}$ decreases. 
The behavior is approximately linear indicating that dispersion and attenuation in the delay line do not play a significant role
in distorting the time correlation. The anti-correlation results from the constant length
of the delay line. Points that lie off this line are consequently spurious and 
can be rejected. One can clearly resolve twelve peaks in the spectrum which correspond to the slits in the mask. 
From the two-dimensional spectrum evident in Fig.~\ref{fig:TCorrel}, 
a one-dimensional spectrum, $\Delta$t$_{Ydown}$ - $\Delta$t$_{Yup}$, is constructed. This spectrum is calibrated using the 2mm center-to-center spacing of the slits. The result is
depicted in Fig.~\ref{fig:YPos}. 
The average width of the central 7 peaks in the
spectrum, $<$$\sigma_{statistical}$$>$, was utilized to determine the spatial resolution. Based on the Gaussian-like nature
of the peaks, the average FWHM was calculated using FWHM=2.35*$\sigma$. Using this
approach the spatial resolution of the detector was determined to be 
520$\mu$m FWHM. An improvement was made by applying 
a Fast Fourier Transform (FFT) filter with a cutoff frequency of 150MHz. This improved the spatial resolution to 488$\mu$m FWHM. The spatial resolution was further improved by selecting events with 70ns 
$<$ ($\Delta$t$_{Ydown}$ + $\Delta$t$_{Yup}$) $<$ 80ns as indicated by the solid lines in Fig.~\ref{fig:TCorrel}. 
With this requirement the spatial resolution improved somewhat to 482$\mu$m with a rejection of 8$\%$ of the events.
Further requirement that the amplitude of the delay line signals exceeded -50mV resulted in the best resolution obtained.  Collectively 
these selection criteria are referred to 
as the ``clean condition'' and resulted in a resolution of 
and 413$\mu$m FWHM. Imposing this condition resulted in a rejection of 66$\%$ of the total events.

\begin{figure}
\begin{center}
\includegraphics[scale=0.30]{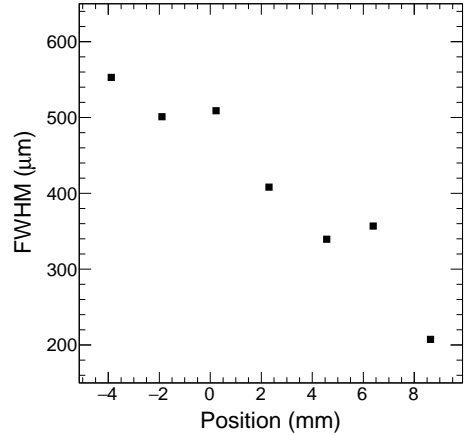}
\caption{Spatial resolution as a function of position for events meeting selection criteria of the the ``clean condition''.} 
\label{fig:FWHMvsYPos}
\end{center}
\end{figure}

The dependence of the measured resolution on position is presented in Fig.~\ref{fig:FWHMvsYPos} for signals which meet the
selection criteria of the ``clean condition''. A clear general trend is 
evident with the resolution decreasing from approximately 550 $\mu$m to approximately 200 $\mu$m over a distance of 
approximately 12 mm. The poorer resolution is associated with the location on the MCP furthest from the foil. This trend was 
qualitatively discernible in Fig.~\ref{fig:YPos}. Relaxing the amplitude requirement results in the same overall trend with 
a slightly larger resolution from 697 to 249 $\mu$m.

\section{Intrinsic Spatial Resolution of E$\times$B MCP Detector}

The spatial resolution measured corresponds to the convolution of the 
intrinsic spatial resolution of the detector with the finite slit width. The measured 
resolution $M(Y')$ is given by:

\begin{equation}
M(Y') = \int{}{} G(Y')I(Y')dY' 
\end{equation}

\begin{figure}
\begin{center}
\includegraphics[scale=0.5]{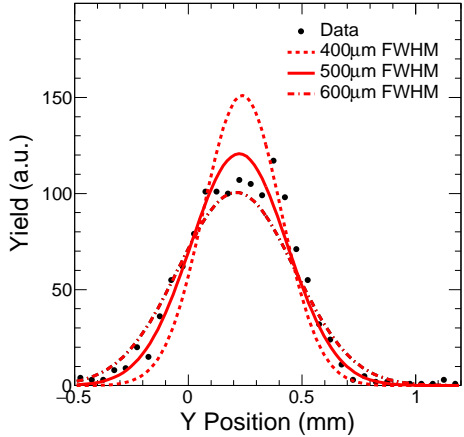}
\caption{Measured one-dimensional position spectrum for the central slit in the mask. Gaussians
with widths of 400$\mu$m, 500$\mu$m, and 600$\mu$m FWHM are depicted for reference.} \label{fig:YPosWidths}
\end{center}
\end{figure}

where $G(Y')$ is taken as a step function with a width of 355$\mu$m to 
represent the slit, and
$I(Y')$ is a Gaussian with the intrinsic width, $\sigma_{intrinsic}$. For a given 
intrinsic width the measured resolution can be calculated. By varying the intrinsic width, the relationship between
intrinsic resolution and measured resolution can be established. This relationship allows one to relate the 
experimentally measured resolution to the intrinsic resolution. In
Fig. \ref{fig:YPosWidths}, superimposed on the experimental data are the predicted resolutions, M(Y'), of 400$\mu$m, 500$\mu$m, and 600$\mu$m FWHM. From this comparison, one can clearly deduce that the measured spatial resolution for this slit is consistent with approximately 
500$\mu$m FWHM. To extract an average intrinsic resolution,
the effect of the finite slit width was de-convoluted from the measured resolution for each of the central 7 peaks. 
The extracted intrinsic resolution of each of the individual peaks was then averaged.
Using this approach an average intrinsic 
resolution of 334$\mu$m FWHM was determined. The measured resolutions for different selection and anaysis criteria 
along with the  intrinsic resolution are summarized in Table~\ref{table:resolution}.

\begin{table}[ht]
\caption{Spatial resolution achieved for different stages in the analysis along with the intrinsic resolution. The 
values shown represent the average over the central seven slits.}
\centering 
\begin{tabular}{l c}
\hline
 & FWHM ($\mu$m) \\ [0,6ex] 
\hline
Raw &  520 \\ 
FFT &  488 \\
FFT + $\Sigma$Delay & 482\\
FFT + $\Sigma$Delay + $>$50mV & 413  \\ 
Intrinsic (FFT + $\Sigma$Delay + $>$50mV ) & 334   \\ [1ex] 
\hline 
\end{tabular}
\label{table:resolution} 
\end{table}


\begin{figure}
\begin{center}
\includegraphics[scale=.35]{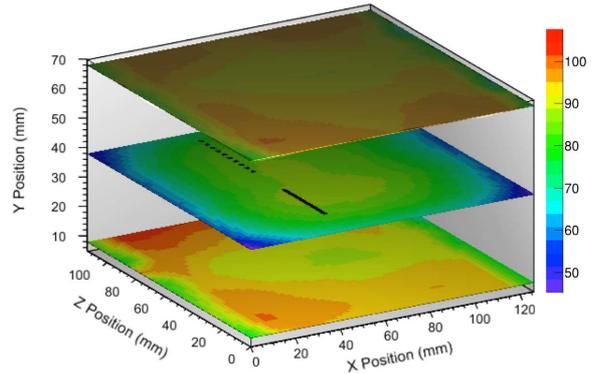}
\caption{(Color online) The magnetic field component, B$_{Y}$, in the XZ plane is presented for the y = 8, 38, 68 mm. The Y-dimension is defined as the principal direction of the magnetic field while the X-dimension is
associated as the direction of the incident ionizing particles. The Z-dimension is perpendicular to both the X and Y dimensions using the right hand rule. The solid and dashed lines shown in the Y=38mm plane indicate the positions of the foil and MCP respectively.} \label{fig:Byfield}
\end{center}
\end{figure}

\section{Simulating the detector resolution}

The significantly larger spatial resolution of 520 $\mu$m obtained with the ExB detector as compared to the 94 $\mu$m \cite{Wiggins17} associated with the 
simple electrostatic arrangement \cite{deSouza12, Wiggins15} indicates that the electron transport from the foil to the MCP dictates the measured resolution.
To understand the electron transport in the
crossed electric and magnetic fields between the secondary-emission foil and the front surface of the MCP detector we simulated the electron trajectories
using the ion trajectory code SIMION \cite{SIMION}.
Performing these calculations required mapping the magnetic field accurately as described below. 

\begin{figure}
\begin{center}
\includegraphics[scale=0.4]{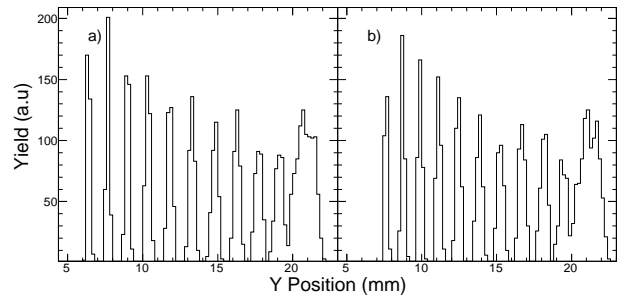}
\caption{SIMION simulations of the position spectrum of electrons generated by $\alpha$ particles passing through slits in the calibration mask. 
The initial kinetic energy and angular distribution of the electrons was assumed to be 3eV and a cone with a 30$^{\circ}$ half angle respectively.
Panel a) A constant magnetic field of B$_Y$ = 90G was used together with
B$_X$ and B$_Z$ = 0G. Panel b) A measured magnetic field was used whose principal axis is presented in Fig.~\ref{fig:Byfield}.} \label{fig:SimYPos}
\end{center}
\end{figure}

\subsection{Mapping the Magnetic Field}

The magnetic field in the active detector volume was measured
using the DC Gaussmeter model GM1-ST \cite{Alpha}. This probe has a manufacturer quoted resolution of 0.1G and an accuracy of 1$\%$ of the measured value. 
This probe was capable of measuring one component
of the magnetic field at a time. The probe was attached to a machined aluminum block and moved in a precise manner to map the
magnetic field in increments of 12.7mm in each dimension. In this manner a two-dimensional plane of one component of the 
magnetic field was produced. By use of precision spacers additional magnetic grid planes were measured. The resulting three-dimensional grid was interpolated to the 1mm level and used for the subsequent simulations. The X-dimension is
defined along the beam axis, the Y-dimension is defined as the principal direction of the magnetic field, and the Z-dimension is perpendicular to both the X and Y dimensions. 
The other components of the magnetic field were measured by rotating the aluminum 
block with the probe attached and repeating the procedure.
The B$_{Y}$ component of the magnetic field is shown in Fig.~\ref{fig:Byfield}. The B$_{Y}$ component of the magnetic field is
shown in the XZ plane for three positions in the Y-dimension, with y=38mm corresponding to the center of the detector. 
As is evident from the figure, the magnetic field in a plane exhibits some asymmetry. We attribute this asymmetry to the 
magnets not being identical and to their placement on the iron plates.
The solid and dashed lines for y=38mm indicate the positions of the foil and MCP respectively. In the active region, the variation of the magnetic field in the principal direction is  approximately $\pm$5G.

\subsection{Simulating the Spatial Resolution of the Detector}

Using the measured magnetic field together with the SIMION-calculated electric field, the trajectory of electrons in the detector was simulated using the program SIMION \cite{SIMION}.
To evaluate the spatial resolution, 100,000 electrons were generated on the masked, aluminized mylar foil and transported through the crossed magnetic and electric fields. An initial kinetic energy of 3eV was assumed for the electrons, consistent with the most 
probable initial electron energy for a similar experimental
setup \cite{Villette00}. 
Although the initial kinetic energy distribution of ejected electrons extends out to 100 eV \cite{Villette00}, it suffices to 
choose 3eV for the initial energy
as the large bias potential applied to the electron emission foil overwhelms the initial kinetic energy chosen.
To begin, we simulated the electron transport in a magnetic field with B$_Y$ = 90G and B$_X$ = B$_Z$ = 0G. Electrons were assumed to be emitted within an initial angular cone with a half angle of 30$^\circ$\cite{Villette00} with one electron emitted for each incident $\alpha$ particle. The simulated position spectrum on the MCP is depicted in Fig.~\ref{fig:SimYPos}a.  Eleven peaks are observed with the peak at Y=21 mm significantly broader than the rest. The average spatial resolution of 
the ten narrower peaks observed in Fig.~\ref{fig:SimYPos}a is 572$\mu$m FWHM. We also investigated the electron transport in the measured magnetic field for the same initial kinetic energy and angular distribution previously used. The simulated position spectrum for the measured magnetic field is depicted in 
Fig.~\ref{fig:SimYPos}b. The results of this magnetic field 
calculation are fairly similar to the previous case. 
The slight difference observed is that the broad peak at Y$>$20 mm 
appears to be splitting into two peaks as evident in Fig.~\ref{fig:SimYPos}b. The average spatial resolution of 
the ten peaks with Y$<$ 20mm is 647$\mu$m FWHM. 
The larger spatial resolution associated with the calculations utilizing the 
measured magnetic field indicate that variations in the field 
degrade the resolution.
Efficiency was defined as the percentage of initial electrons originating from the slits that subsequently reach the multi-strip anode, which was determined to be $\sim$ 88$\%$ for both the measured and constant field simulations. We also investigated the position sensitivity in the Z-dimension using the same approach that was used in the Y-dimension. The position sensitivity in this dimension was determined to be $\sim$ 7mm.

\begin{figure}
\begin{center}
\includegraphics[scale=0.35]{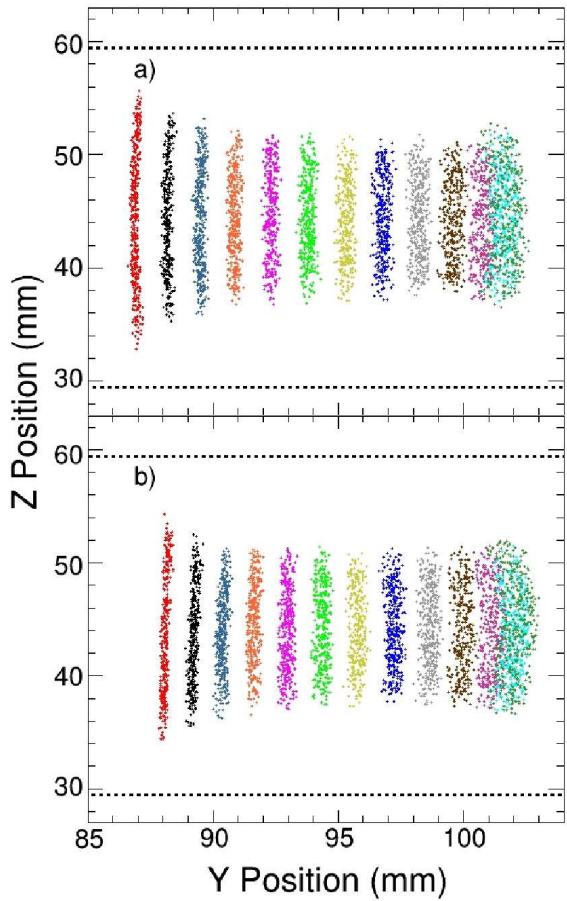}
\caption{(Color online) (a) Slit image predicted by SIMION at the anode position for the case of B$_Y$ = 90G and B$_X$ = B$_Z$ = 0G. (b) Slit image predicted by SIMION at the anode position for the case of the measured magnetic field.} \label{fig:YZSIMION}
\end{center}
\end{figure}

We investigated why the spectrum evident in Fig.~\ref{fig:SimYPos}b only exhibits 12 peaks. For the $\alpha$-particles incident on the 13 slits, the final YZ positions are shown in
Fig.~\ref{fig:YZSIMION}. In the case of B$_Y$ = 90G and B$_X$ = B$_Z$ = 0G, presented in Fig.~\ref{fig:YZSIMION}a, the image of the slits is observed as 
twelve vertical stripes. For reference, the area of the multi-strip anode is depicted by the dashed lines. As one moves to larger Y position, the width of each stripe increases resulting in poorer resolution. For the two  
slits at the largest value of Y, the slits are unresolved from each other. Thus the trend of resolution with position observed in Fig.~\ref{fig:FWHMvsYPos} is reproduced. Two additional features can be noted  
in Fig.~\ref{fig:YZSIMION}b. 
The vertical extent of the central stripes is somewhat smaller than that of the edge stripes indicating a contribution of 
focusing in the image by the field. In addition,   
the rightmost stripe exhibits a curvature not observed for the other stripes.
Given the 7mm resolution in Z, by replacing the multi-strip anode with a cross-strip anode \cite{Siegmund09} 
a two-dimensional position measurement could be implemented.

\section{Conclusion}

An E$\times$B MCP detector with position-sensitivity in 1-dimension has been realized. Position-sensitivity was achieved by utilizing a MCP coupled to a multi-strip anode with delay line readout.
Signals arriving at either end of the 
delay line were digitized by high speed digitizers and subsequently analyzed. 
To measure the position-sensitivity, a mask was inserted and the detector was exposed to $\alpha$-particles from an $^{241}$Am source. While the simplest analysis provided a measured spatial resolution of 520$\mu$m FWHM, use of digital signal processing techniques along with use of signal selection criteria
improved the spatial resolution to 413$\mu$m. 
This measured
resolution of 413$\mu$m FWHM corresponds to an intrinsic resolution of 334$\mu$m FWHM. To understand the measured resolution, the magnetic field was mapped, and the trajectories of ejected electrons were calculated using the program SIMION. For a constant magnetic field of B$_Y$ = 90G and B$_X$ = B$_Z$ = 0G, simulations predict a spatial resolution of
572$\mu$m 
FWHM. Use of the measured magnetic field results in a spatial resolution of 647 $\mu$m FWHM. This approximate agreement of the simulation with the measured resolution suggests that the primary factors that influence the resolution are understood. 
The primary factor that dictates the measured resolution is the electron transport from the foil to the MCP.
Although the present design provides good position sensitivity over a limited region in one-dimension, this initial development of a compact, 
high-rate position-sensitive E$\times$B detector is promising.

\section{Acknowledgments}

We gratefully acknowledge the technical support provided by the personnel in the 
Mechanical Instrument Services and Electronic Instrument Services (EIS) at the Department of Chemistry,
Indiana University. In particular, we gratefully acknowledge A. Alexander of EIS for the design of the multi-strip anode and delay printed circuit boards.
We thank Mr. Luis Morales (Notre Dame University) for providing additional magnetic field calculations which aided our understanding 
of the detector performance.
This research is based upon work  supported by the U.S. Department of Energy under Award Number FG02-88ER-40404, the
National Nuclear Security Administration under Award Number DE-NA0002012, and the National Science Foundation under Grant Number 1342962.

\bibliographystyle{elsarticle-num-names}

\end{document}